\documentclass[
    ,final            
  ]
  {aipproc}

\layoutstyle{6x9}
\usepackage{natbib}
\bibpunct{(}{)}{;}{a}{}{,}
\newcommand{\aap}{A\&A}

\newcommand{\apj}{ApJ}

\newcommand{\mnras}{MNRAS\ }

\newcommand{\araa}{ARA\&A\ }

\def\mrk{Mrk~279}
\def\ch{{\it Chandra}}
\def\hst{HST-STIS}

\def\Halpha{\ifmmode {\rm H}\alpha \else H$\alpha$\fi}
\def\Hbeta{\ifmmode {\rm H}\beta \else H$\beta$\fi}
\def\Hgamma{\ifmmode {\rm H}\gamma \else H$\gamma$\fi}
\def\Hdelta{\ifmmode {\rm H}\delta \else H$\delta$\fi}
\def\Lya{\ifmmode {\rm Ly}\alpha \else Ly$\alpha$\fi}
\def\Lyb{\ifmmode {\rm Ly}\beta \else Ly$\beta$\fi}
\def\Lyg{\ifmmode {\rm Ly}\beta \else Ly$\gamma$\fi}

\def\ciii{\ifmmode {\rm C}\,{\sc iii} \else C\,{\sc iii}\fi}
\def\civ{\ifmmode {\rm C}\,{\sc iv} \else C\,{\sc iv}\fi}
\def\cv{\ifmmode {\rm C}\,{\sc v} \else C\,{\sc v}\fi}
\def\cvi{\ifmmode {\rm C}\,{\sc vi} \else C\,{\sc vi}\fi}

\def\nvi{N\,{\sc vi}}
\def\nvii{N\,{\sc vii}}
\def\oi{O\,{\sc i}}
\def\oii{O\,{\sc ii}}

\def\o5007{[O\,{\sc iii}]\,$\lambda5007$}

\def\ovii{O\,{\sc vii}}
\def\oviii{O\,{\sc viii}}

\def\neix{Ne\,{\sc ix}}
\def\nex{Ne\,{\sc x}}

\begin{document}

\title{Chandra-LETGS observation of Mrk~279: modeling the AGN internal structure}


\author{E.~Costantini}{
  address={SRON, Sorbonnelaan, 2, 8534CA Utrecht, The Netherlands}
  ,altaddress={Astronomical Institute, Utrecht University, P.O. Box 80000, 3508TA Utrecht, The Netherlands} 
}

\author{J.S.~Kaastra}{
  address={SRON, Sorbonnelaan, 2, 8534CA Utrecht, The Netherlands}
}

\author{K.C.~Steenbrugge}{
  address={SRON, Sorbonnelaan, 2, 8534CA Utrecht, The Netherlands}
}
\author{N.~Arav}{
  address={CASA, University of Colorado, 389 UCB, Boulder, CO, USA}
}
\author{J.R.~Gabel}{
  address={CASA, University of Colorado, 389 UCB, Boulder, CO, USA}
}
\author{G.~Kriss}{
  address={STSI, 3700 San Martin Drive, Baltimore, MD 21218, USA}
}

\begin{abstract}
We present the \ch-LETGS analysis of \mrk, a bright Seyfert~1 galaxy. The spectrum shows a variety of features arising from different
physical environments in the vicinity of the black hole. We detect three absorption components, located at the redshift of the source. One
of them is likely to arise from the host galaxy. An additional component due to absorption by a collisionally ionized gas at z=0 is also
observed. Such an absorption can be produced in the outskirts of our Galaxy or in the local group.  
The emission spectrum is rich in narrow and broad emission features. We tested
the hypothesis that the broad emission lines originate in the Broad Line Region (BLR), as studied in the UV band. 
We find that at least 2 components with different ionization parameters are necessary to
simultaneously interpret the broad emission line measured in the UV (by \hst\ and FUSE) and in the X-rays.

\end{abstract}

\maketitle


\section{Introduction}
The high spectral resolution now available in the X-ray band led to a deeper understanding of the physical processes occurring in active
galaxies. \mrk\ is a type~1 Seyfert galaxy. For these objects, according to the standard model \citep[e.g. ][]{antonucci}, 
our line of sight should intercept an outflowing, highly ionized gas. It is now recognized that 
many components, differing in ionization level, column density, and outflowing velocity, are actually constituting this outflow
\citep[e.g.][]{jelle00,kaspi01,katrien04} and that the UV and the X-ray band share at least some of these components
\citep[e.g.][]{katrien04,krongold}. 
Yet, the geometry, the distance, and therefore the density of the X-ray outflow is not yet clear \citep[e.g.][]{behar03,katrien04,krongold}. 
The observation angle for Seyfert~1 objects allows to investigate the innermost parts of the AGN, from the broad line region 
\citep[BLR, see e.g. ][]{baldwin}, at a distance of few light days from the black hole, down to the accretion disk. 
The bulk of
the X-rays are produced here \citep[e.g. ][]{nandra}. The BLR is likely to produce X-ray radiation, 
but up to now this radiation was never convincingly observed in X-ray spectra \citep{katrien04,ogle04}.          
   
\section{Results}
\begin{figure}
   \includegraphics[height=.38\textheight,angle=90]{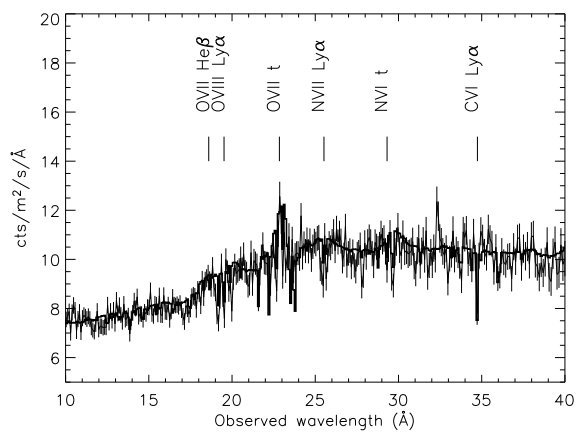}
  \caption{\label{f:best_fit}Best fit model for the LETGS data of \mrk. The labels refer to the broad lines added in the model.}
\end{figure}

We processed the \ch-LETGS data of \mrk\ following the approach described in \citep{jelle02}. The residuals to a basic continuum model show
significant excess emission, especially in the 18-24 \AA\ wavelength range, where the main oxygen line features are (\ovii\ 
triplet and \oviii\  \Lya). We interpret these emission features as broad lines, possibly arising from the BLR. The
best fit shows that the excess can be modeled by Gaussian profiles for \ovii, \oviii, \nvii, and \cvi. 
Evidence of weaker broad excesses is found for \nvi\ and \cv\ (Fig.~\ref{f:best_fit}) and possibly neon (see next
paragraph for the modeling). 
The inclusion of broad emission lines is required to properly fit the absorbed spectrum.
The best fit requires indeed absorption by two components of ionized gas. For the lower ionized component we
find: ${\rm N_H}\sim 1.6\times10^{22}$cm$^{-2}$, and log$\xi\sim$0.57, while for the high ionization component the parameters are: 
N$_{\rm H}\sim 4\times10^{22}$cm$^{-2}$ and log$\xi\sim$2.5. The two absorbers seem to belong to systems with different 
outflow velocity, -220$_{-90}^{+50}$ and -570$_{-70}^{+100}$ km\,s$^{-1}$ for the low and high ionization gas, respectively. 
Another component, whose main features are \oi\ and \oii\ has been modeled by a collisionally ionized gas at a temperature $t\sim 8$ 
eV. \mrk\ shows also ionized absorption taking place at z=0 \citep{jelle05}. In the LETGS band 
the deepest features of this collisionally ionized component are \ovii\ and \cvi.     

\subsection{The Broad Line Region}

\begin{figure}
   \includegraphics[height=.38\textheight,angle=90]{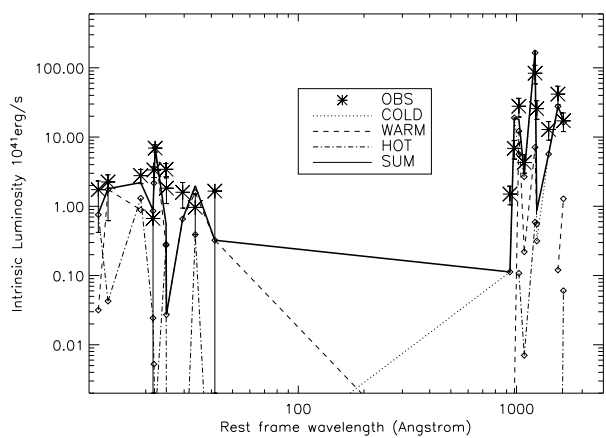}
 \caption{\label{f:cloudy}Cloudy prediction of the ionization state of the BLR emitting clouds (see text).}
\end{figure}

Thanks to the simultaneous FUSE and HST observation \citep{arav}, we could compare the X-ray line luminosity with the ones measured for
the UV broad emission lines. We used 
Cloudy \citep{ferland} for the modeling. The input ionizing spectral continuum was evaluated directly from the HST, FUSE and \ch\ data. 
We chose a grid of physical parameters for the BLR clouds and then chose the best fit to our data through $\chi^2$
minimization. In Fig.~\ref{f:cloudy} we display the intrinsic integrated luminosity measured for the X-ray and UV lines (star points) 
{\it vs} the rest frame wavelength of
the lines, together with our best fit. 
Two ionization components for the emitting medium are certainly needed in order to explain the UV and the lower
ionization X-ray lines (labeled as ``cold" and ``warm" in Fig.~\ref{f:cloudy}). 
This result is based mainly on the more precise UV measurements. 
The higher ionization X-ray lines, such as \cvi, \nvii, \oviii, \neix, and \nex\ may require the contribution of a higher
ionization component (``hot" in Fig.\ref{f:cloudy}). 
The ionization parameters of these components are roughly log$\xi\sim$ 
-1.5, 1.6 and 3.6 for the ``cold", ``warm", and ``hot" components, respectively. 
The distance from the ionizing source of the emitting cloud is around $10^{16-17}$ cm, and the density ranges between $10^{9-12}$cm$^{-3}$. 
The ionization parameter is given by: $\xi=L/nr^2$.
Here only the ionizing luminosity $L$ is directly measured. Density effects on the emission may help in determining the
parameters in the $nr^2$ term. However, as a caveat, the $r$ and $n$ values are still coupled.


\bibliographystyle{aipprocl} 




\end{document}